\documentclass[aps,twocolumn,floats,prd,nofootinbib,superscriptaddress]
{revtex4-2}

\usepackage[utf8]{inputenc}

\usepackage{graphicx,amsmath,amsfonts,amssymb,slashed}
\usepackage{bbold,wasysym}
\usepackage{hyperref}

\usepackage{amsmath,amsfonts,bbm,euscript,hyperref}

\usepackage{xcolor}
\usepackage{color}
\usepackage{hyperref}

\hypersetup{colorlinks, citecolor=blue, linkcolor=red, urlcolor=blue}

\usepackage{graphicx,capt-of}
\usepackage{multirow}
\usepackage{footnote}
\usepackage{ctable}

\usepackage{amsmath}
\usepackage{lipsum}

\usepackage[english]{babel}
\usepackage{blindtext}

\newcommand{\bs}[1]{\boldsymbol{#1}}

\definecolor{bguorange}{RGB}{244, 119, 33}

\definecolor{mygreen}{rgb}{0.01, 0.75, 0.24}

\newcommand\mnras{MNRAS}             

\begin{document}

\title{Constraining Primordial Magnetic Fields with Line-Intensity Mapping}

\author{Tal Adi}
\email{talabadi@post.bgu.ac.il}
\affiliation{Department of Physics, Ben-Gurion University of the Negev, Be'er Sheva 84105, Israel}

\author{Sarah Libanore}
\email{libanore@bgu.ac.il}
\affiliation{Department of Physics, Ben-Gurion University of the Negev, Be'er Sheva 84105, Israel}

\author{Hector Afonso G.\ Cruz}
\email{hcruz2@jhu.edu}
\affiliation{William H. Miller III Department of Physics and Astronomy, Johns Hopkins University, 3400 N. Charles Street, Baltimore, Maryland, 21218, USA}

\author{Ely D. Kovetz}
\email{kovetz@bgu.ac.il}
\affiliation{Department of Physics, Ben-Gurion University of the Negev, Be'er Sheva 84105, Israel}

\begin{abstract} 
Primordial magnetic fields (PMFs) offer a compelling explanation for the origin of
observed magnetic fields, especially on extragalactic scales. Such PMFs give rise to excess of power in small scale matter perturbations that could strongly influence structure formation. We study the impact of the magnetically enhanced matter power spectrum on the signal that will be observed by line-intensity mapping (LIM) surveys targeting carbon monoxide (CO) emission from star-forming galaxies at high redshifts.
Specifically, the voxel intensity distribution of intensity maps provides access to small-scale information, which makes it highly sensitive to signatures of PMFs on matter overdensities. We present forecasts for future LIM CO surveys, finding that they can constrain PMF strength as small as $B_{\rm 1Mpc}\sim0.006-1\,{\rm nG}$, depending on the magnetic spectral index and the targeted redshifts.
\end{abstract}
                            
\maketitle

\section{Introduction}
Magnetic fields are a pervasive feature of the Universe, existing in a wide range of environments from planets~\cite{Stevenson:2010} and stars~\cite{Hathaway:2010fk}, to galaxies~\cite{2001SSRv...99..243B,2013pss5.book..641B} and galaxy clusters~\cite{Clarke:2000bz,Govoni:2004as,2005A&A...434...67V}, at scales ranging between $\sim1\,$Gauss to $\sim10^{-6}\,\mathrm{Gauss}$. There is also observational evidence for $\sim10^{-7}\,$Gauss magnetic fields, which are coherent on scales up to tens of kpc inside and nearby high redshift galaxies~\cite{Fletcher:2010wt, 2008Natur.454..302B, MaoEtAl2017}, as well as in the interstellar medium~\cite{2017ARA&A..55..111H}. While magnetic fields on all these scales play a crucial role in astrophysical processes and can have significant effects on the behavior of matter and radiation, their origin remains a mystery. 
The dynamo process~\cite{1955ApJ...122..293P}, which amplifies a seed field, is believed to be the mechanism responsible for generating the observed magnetic fields on different scales~\cite{2013MNRAS.429.2469B, 2018MNRAS.475L..72S, 1955ApJ...122..293P}. Various models have been proposed for the generation of  such seeds on small~\cite{1950ZNatA...5...65B, 2002RvMP...74..775W, 2005ApJ...633..941H, 2018MNRAS.479..315S} and large~\cite{1994MNRAS.271L..15S,2006AstHe..99..568I, 2008Sci...320..909R, 2013PhRvL.111e1303N, SchoberEtAl2013} scales. 

Meanwhile, recent observational evidence based on blazar emissions suggest that weak $\sim10^{-16}\,$Gauss magnetic fields, coherent on Mpc scales, could likewise exist in the intergalactic medium (IGM)~\cite{2010Sci...328...73N, Tavecchio:2010mk, Taylor:2011bn}. Yet, the existence of these fields is difficult to explain with purely astrophysical processes in the late Universe.
An intriguing explanation for magnetic fields on such scales could be their production at an early phase of the Universe. The generation of such primordial magnetic fields (PMFs) could be achieved  via numerous processes in the early Universe, e.g., inflation~\cite{PhysRevD.37.2743,Turner:1987bw,Ratra:1991bn} or phase transitions~\cite{Hogan:1983zz,Vachaspati:1991nm,Grasso:1997nx}; for reviews see Refs.~\cite{Grasso:2000wj, Kandus:2010nw, 2016RPPh...79g6901S}. 
The presence of magnetic fields early on could leave a signature on different cosmological and astrophysical observables; for example, magnetized gas may affect the polarization of the cosmic microwave background (CMB) by inducing Faraday rotation~\cite{scoccola04, kosowsky05, kahniashvili09, yamazaki10, pogosian11, Planck:2015zrl}, just as dissipation of PMFs can heat the IGM, thus affecting structure formation~\cite{Sethi:2004pe, Sethi:2008eq, Schleicher:2008aa}.

Perhaps the most prominent signature of PMFs is on the matter power spectrum. Prior to recombination, PMFs evolve according to the laws of magneto-hydrodynamics (MHD)~\cite{Jedamzik:1996wp,Subramanian:1997gi}, as they interact with the ionized baryon plasma. Consequently, the ionized plasma is subjected to Lorentz pressure, which induces additional matter density fluctuations~\cite{wasserman1978,coles92,Kim:1994zh,Ralegankar:2023pyx}. These tend to cause an excess at small scales $k\gtrsim1\,\mathrm{Mpc}^{-1}$ in the matter power spectrum, which in turn affects observables that trace structure formation. There have been several attempts to detect or place constraints on PMFs through observations and experiments~\cite{Shaw:2010ea, Kahniashvili:2012dy, Fedeli:2012rr, Pandey:2012ss, Pandey:2014vga, Planck:2015zrl, Sanati:2020oay}. However, overall the constraints on the strength of PMFs are still not very tight, mainly due to the low sensitivity to information on small scales of the various experiments. We consider, for the first time, a revised expression for the magnetic contribution to the matter power spectrum that is smaller by a factor of $(4\pi)^2$ from previous studies that made use of this formalism, as described in Ref.~\cite{Adi:2023doe}.

A promising probe of structure formation in the Universe is line intensity mapping (LIM), which measures the integrated spectral line emission from galaxies and the IGM~\cite{Kovetz:2017agg, Breysse:2016szq, Bernal:2022jap}. Depending on the observed frequency range, many lines can be observed at different redshifts, probing different phases the IGM and star-forming regions. There are ongoing and planned experiments~\cite{Cleary:2021dsp, CCAT-prime, Ade:2019ril, TIM, Dore:2014cca, Hill:2008mv} targeting neutral and ionized hydrogen lines, e.g.,~21cm spin-flip of neutral hydrogen, Lyman-$\alpha$, H$\alpha$, H$\beta$~\cite{Lidz:2011dx, Breysse:2014uia, Li:2015gqa, Pullen:2012su, Padmanabhan:2017ate,Silva:2014ira, Pullen:2017ogs,2017ApJ...835..273G, 2018MNRAS.475.1587S,Pullen:2013dir, 2013ApJ...763..132S,Bernal:2019gfq}, as well as atomic and molecular lines produced in star forming regions, e.g.~CO, CII, OIII; see Ref.~\cite{Bernal:2022jap} for a recent review. Among these, the   rotational carbon-monoxide (CO) transitions are of particular interest~\cite{Righi:2008br, Lidz:2011dx, Pullen:2012su, Breysse:2014uia, Breysse:2015saa}, and in particular CO(1-0), as its line-foreground contamination is relatively low~\cite{Chung:2017uot,COMAP:2021nrp} and it can be easily observed from galaxies at $z\sim 3-7$.

It was recently shown in Ref.~\cite{Libanore:2022ntl} that LIM surveys can probe the matter distribution on both small and large scales through the one-point function of the intensity fluctuations measured in their maps. This is commonly referred to as the voxel intensity distribution (VID)~\cite{Breysse:2015saa,Breysse:2016szq}. 
Unlike higher order statistics, which suffer from limited angular resolution, the VID is sensitive to the total emission from galaxies formed in both small and massive halos (see e.g., Ref.~\cite{Sato-Polito:2022wiq}). 
This suggests that the VID of a LIM survey may serve as a probe for magnetically induced matter perturbations on small scales.

In this work, we study the impact of the magnetically induced matter power spectrum contribution due to PMFs, on the VID of the CO(1-0) line intensity maps. We consider a PMF power spectrum in the form of a power law, parameterized by $B_{\rm 1Mpc}$ and $n_B$, which specify its strength and spectral index, as described below. We find that LIM surveys will be able to place stringent constraints on PMFs, in the range of $B_{\rm 1Mpc}\sim 0.04-1\,{\rm nG}$, determined by $n_B$ and the specifications of the survey.

The paper is organized as follows. In Section~\ref{sec:methods} we describe the theory our work relies on, divided into two parts: Sec.~\ref{sec:pmf_method}, in which we review the magnetically induced matter power spectrum, and Sec.~\ref{sec:vid_method}, in which we highlight the relevant details in the VID formalism (further detail on both topics can be found in the appendices). In Section~\ref{sec:analysis}, we describe the implementation and set-up for our analysis. In Section~\ref{sec:results} we present and discuss our results for the VID constraining power on PMFs. We conclude our work in Section~\ref{sec:conclusions}.

\section{Methods}
\label{sec:methods}

\subsection{Magnetically Induced Matter Perturbations}
\label{sec:pmf_method}

We consider a phenomenological model for the PMFs and their impact on the evolution of matter perturbations. The formalism we use was first laid out in Ref.~\cite{wasserman1978} and used in previous works~\cite{Kim:1994zh,2003JApA...24...51G, Shaw:2010ea, Pandey:2012ss,Fedeli:2012rr}.
We summarize the main points of the derivation in Appendix~\ref{app:PMF} for completeness.
\subsubsection{Modelling of the PMFs}
Assuming that PMFs were generated via some process during the early Universe (see e.g., Refs.~\cite{Grasso:2000wj, Kandus:2010nw, 2016RPPh...79g6901S} for review), so that they are initially isotropic and homogeneous, the 
two-point function of the magnetic field's Fourier transform $\tilde{\bs{B}}(\bs{k}, t) = \int dx\,\bs{B}(\bs{x},t)e^{-i\bs{k}\cdot\bs{x}}$, can be written as~\cite{1959flme.book.....L,1967PhFl...10..859K}
\begin{equation}
    \left\langle\tilde{B}_i(\bs{k}, t) \tilde{B}_j^*\left(\bs{q}, t\right)\right\rangle=\frac{(2 \pi)^3}{2} \delta_D\left(\bs{k}-\bs{q}\right)\mathrm{P}_{ij}(\bs{k}) P_B(k, t)\label{eq:two-point-B}
\end{equation}
where $\mathrm{P}_{ij}(\bs{k})=\delta_{ij}-k_iq_j/k^2$ and $P_B(k,t)$ is the PMFs power spectrum. We assume a phenomenological model in which the power spectrum takes the form of a power law (e.g.~\cite{Kim:1994zh})
\begin{equation}
    P_B(k,t) = A_B(t) k^{n_B},\quad \mathrm{ for }\quad k_\mathrm{min}<k<k_\mathrm{max},\label{eq:P_B}
\end{equation}
with free parameters $A_B$ and $n_B$ that determine the strength and spectral index. In practice, it is common to characterize the magnetic field in terms of its physical strength smoothed over a characteristic scale $\lambda$, which we take to be $\lambda=1\,$Mpc,
\begin{align}
    B_{\lambda}^2(t) &= \int_0^{\infty} \frac{ dk\, k^2}{2\pi^2} P_B(k,t)e^{-k^2\lambda^2}\nonumber\\
    &= \frac{A_B(t)}{(2\pi)^2}\frac{\Gamma\left[(n_B+3)/2\right]}{\lambda^{n_B+3}}. \label{eq:sigma_B}
\end{align}
It is also common to consider only $n_B>-3$, due to infrared divergences at lower values of the spectral index. 
The scale limits in Eq.~\eqref{eq:P_B} are determined from the scales in which PMFs are coherent. Since these fields are coherent to very large scales, the value of $k_{\rm min}$ can be taken to be the order of the Hubble radius. On the other hand, the value of $k_{\rm max}$ is determined by the scale beyond which magnetic fields dissipate, as we discuss below.
\subsubsection{Magnetic Fields Damping}
Before recombination, 
the magnetic fields perturbations, which are mainly frozen in the baryon-photon plasma, dissipate on all scales smaller than the radiation diffusion length~\cite{Jedamzik:1996wp,Subramanian:1997gi}. As a result, the magnetic field modes $\tilde{\bs{B}}(\bs{k},t)$ gain a damping factor approximated by $\exp\left[{-k^2/k_A^2}\right]$, where the Alfvén wave number $k_A$ is determined by
\begin{equation}\label{eq:kAlfven}
    \frac{1}{k_A^2} = \int_0^{t_\mathrm{rec}}\frac{v_A^2\tau_c}{a^2(t)}dt,
\end{equation}
where $1/\tau_c = cn_e\sigma_T$ is the Thomson scattering rate, $t_\mathrm{rec}$ is the time of recombination and $v_A$ is the characteristic Alfvén velocity. This velocity is determined from the pressure and energy density of the plasma and the magnetic field, and can be written as
\begin{equation}\label{eq:vAlfven}
    v_A = \frac{B_{\lambda_A}(t)}{\sqrt{\mu_0 \left[\rho(t) + p(t) \right]}} = \frac{B_{\lambda_A}(t)}{\sqrt{\mu_0 \left[\bar{\rho}_b(t)+\frac{4}{3}\bar{\rho}_r(t)\right]}},
\end{equation}
where $\rho$ and $p$ are the energy density and pressure of the plasma, $\bar{\rho}_{b,r}$ denotes baryon and radiation background densities respectively, and $\lambda_A=2\pi/k_A$. Therefore, the damped PMFs power spectrum in Eq.~\eqref{eq:P_B} becomes
\begin{equation}
    P_B(k,t) = A_B(t) k^{n_B} e^{-2k^2/k_A^2},\label{eq:P_B-damped}
\end{equation}
where $k_A$ plays the role of $k_\mathrm{max}$, with an exponential tail instead of a sharp cutoff.
\subsubsection{Impact on the Linear Matter Power Spectrum}
Due to the presence of magnetic fields prior to recombination, the electrically conducting fluid is subjected to a Lorentz pressure that acts as an additional source of matter density fluctuations~\cite{wasserman1978,Kim:1994zh}. However, on small enough scales, the magnetic pressure gradients counteract the gravitational pull \cite{Subramanian:1997gi}, leading to a \emph{magnetic Jeans scale} beyond which the magnetically induced perturbations do not grow~\cite{Kim:1994zh,Sethi:2004pe}. This is given by
\begin{equation}
    \lambda_J = \frac{2\pi}{k_J} = \left[ \frac{16\pi}{25}\frac{B_{\lambda,0}^2}{\mu_0G\bar{\rho}_{m,0}\bar{\rho}_{b,0}}\lambda^{3+n_B} \right]^{1/\left(5+n_B\right)},\label{eq:jeans-scale}
\end{equation}
where $G$ is Newton's constant and $\bar{\rho}_{m,0}$ and $\bar{\rho}_{b,0}$ are today's background matter and baryon energy densities. Here we used the temporal evolution of the magnetic fields, which follows $\bs{B}(\bs{x},t) = \bs{B}_0/a^2$ on scales much larger than the magnetic Jeans scale; for details see Appendix~\ref{app:PMF}. Moreover, magnetic fields dissipate on scales smaller than $\lambda_J$ due to ambipolar diffusion and decaying turbulence~\cite{Sethi:2004pe, Sethi:2008eq, Kahniashvili:2010wm} that heat the IGM. However, at a linear level, these scales can be neglected.

In general, the linear matter power spectrum in the presence of PMFs may contain contributions from the cross correlation term between the standard and magnetically induced matter density contrasts, depending on the PMF generation process \cite{Yamazaki:2006mi}. However, since there is no prior information on such correlation, for simplicity we assume that the fields are uncorrelated. 
Thus, the total matter power spectrum for $k<k_J=2\pi/\lambda_J$, is of the form
\begin{equation}
    P_m(k,t) = D_+^2(t)P_\mathrm{lin}(k) + M^2(t)\Pi(k),\label{eq:Pm}
\end{equation}
 where $D_+$ and $M$ are the standard and magnetic growth functions, respectively, and $\Pi(k)$ is calculated according to Eq.~\eqref{eq:PiPower-general}. Using Eq.~\eqref{eq:two-point-B}, this yields\footnote{Note that, in addition to the $1/\mu_0^2$, there is a numerical factor of $1/(4\pi)^2$ that was missed in all previous work we know of, as shown in Ref.~\cite{Adi:4pi}.}~\cite{Kim:1994zh}
\begin{align}
    \Pi(k) = \frac{\alpha^2}{(4\pi)^2} \int  d q \int~d &\mu \frac{P_{B,0}(q) P_{B,0}\left(\left| \bs{k}-\bs{q} \right|\right)}{\left| \bs{k}-\bs{q} \right|^2} \nonumber\\
    \times &  \mathcal{F}(\bs{k},\bs{q}),
\label{eqn:PiPower} 
\end{align}
with $\alpha = f_b/\left(\mu_0 \bar{\rho}_{b, 0}\right)$, $\mu = \hat{\bs{k}}\cdot\hat{\bs{q}}$ and
\begin{equation}
    \mathcal{F}(\bs{k},\bs{q}) = 2 k^5 q^3 \mu+k^4 q^4\left(1-5 \mu^2\right)+2 k^3 q^5 \mu^3.
\end{equation}
The evolution of $M(t)$ is determined according to 
\begin{equation}
    \ddot{M}(t)+2H(t)\dot{M}(t)-4\pi G\bar{\rho}_{m,0}\frac{M(t)}{a^3(t)} = \frac{1}{a^3(t)},\label{eq:M-evo}
\end{equation}
where $H(t)$ is the Hubble parameter, $a(t)$ the scale factor, and assuming null initial conditions at recombination: $M(t_\mathrm{rec})=\dot{M}(t_\mathrm{rec})=0$.

\begin{figure}[ht!]
    \centering
    \includegraphics[width=\columnwidth]{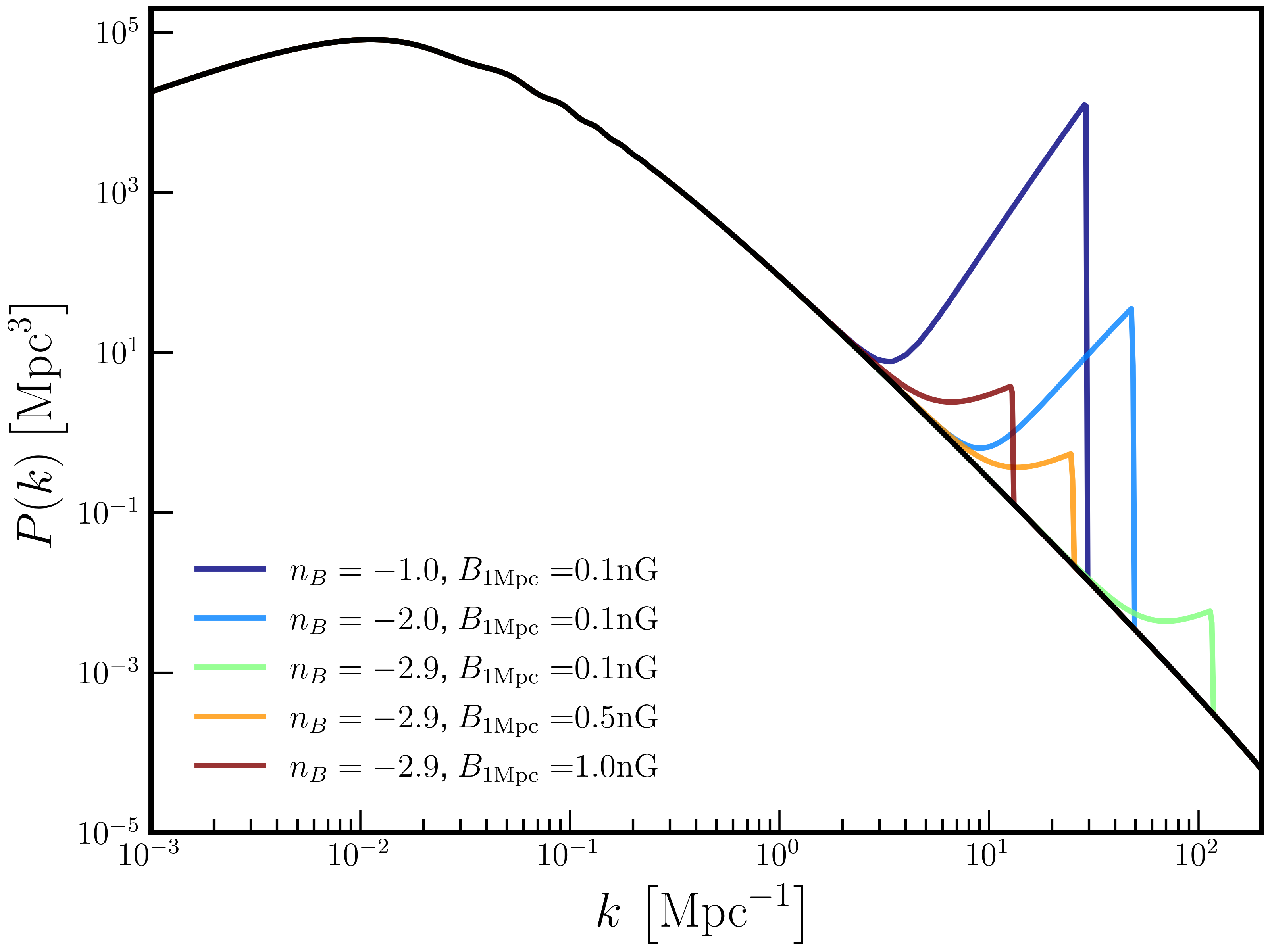}
    \caption{The linear matter power spectrum at $z=0$, as described in Eq.~\eqref{eq:Pm}. We present the resulting spectrum for different values of the PMFs strength and spectral index, as well as the $\Lambda$CDM scenario, in black, which we calculate using \texttt{CLASS} code\footnote{\url{https://github.com/lesgourg/class_public}}~\cite{Blas:2011rf}.}
    \label{fig:pmf_vary}
\end{figure}

\subsection{VID Sensitivity to the Matter Power Spectrum}
\label{sec:vid_method}

We follow Refs.~\cite{Breysse:2014uia, Breysse:2016szq, Bernal:2020lkd, Libanore:2022ntl} and model the VID as the histogram of the number of voxels for which the measured brightness temperature is found within $\Delta T_i$,
\begin{equation}
    B_{i} = N_{\mathrm{vox}}\int_{T_{i}}^{T_{i}+\Delta T_{i}}\mathcal{P}_{\mathrm{tot}}\left(T\right)dT \label{eq:Bivid},
\end{equation}
where $N_\mathrm{vox}$ is the total number of voxels in the survey and $\mathcal{P}_\mathrm{tot}(T)$ is the probability of observing a voxel with a brightness temperature $T$, as we review in Appendix~\ref{app:VidModeling}. In practice, $\mathcal{P}_\mathrm{tot}(T)$ depends on the probability $\mathcal{P}_1(T)\propto dn/dL$ of a single source having brightness temperature $T$, where $dn/dL$ is the luminosity function; and on $\mathcal{P}_s(N_s)$, the probability to find $N_s$ sources in a voxel. Both of these quantities depend on cosmology and astrophysics; here, we only note the quantities affected directly by the modified matter power spectrum, according to Eq.~\eqref{eq:Pm}. The interested reader can refer to Appendix~\ref{app:Vid} or Refs.~\cite{Breysse:2014uia, Breysse:2016szq, Bernal:2020lkd, Libanore:2022ntl} for the full derivation. 

The matter power spectrum enters in $\mathcal{P}_\mathrm{tot}(T)$ in two ways, both explicitly, as in the computation of the source distribution $P_s(N_s)$ described in Appendix~\ref{app:VidSourceDist}, and implicitly, sourcing the halo mass function. 
Once a specific line is chosen for observations,\footnote{Following Ref.~\cite{Libanore:2022ntl}, we rely on surveys targeting the rotational CO $(1\to 0)$ transition sourced by star forming molecular gas clouds at $z \sim 6$. This line is expected to have small line-foreground contamination~\cite{Chung:2017uot,COMAP:2021nrp}, which allows us to neglect, at first stage, the presence of interlopers.} the halo mass function is used to evaluate the luminosity function~\cite{Libanore:2022ntl},
\begin{equation}\label{eq:dndl}
\begin{aligned}
&   \frac{dn}{dL_{\rm CO}}(L) = \int_{M^{\rm min}}^{M^{\rm max}}dM\,\frac{ e^{-L_{\rm cut}/L}}{L\sqrt{2\pi}\sigma^2_{\rm TOT}}\, 
    \frac{dn}{dM}\\
 & \times \exp\biggl[-\frac{\bigl(\log L - \log(L_\mathrm{CO}(M))+\sigma^2_{\rm TOT}/2\bigr)^2}{2\,\sigma^2_{\rm TOT}}\biggr].
\end{aligned}
\end{equation}
Here $L$ is the luminosity of the source, $M$ is the mass of the host dark matter (DM) halo, $L_\mathrm{cut}$ is a luminosity exponential cutoff scale, $L_\mathrm{CO}$ is the CO line emission luminosity, and $\sigma_\mathrm{TOT}$ encapsulates the lognormal scatter in the relation between $L_\mathrm{CO}$ and halo mass $M$, see Appendix~\ref{app:VidLCO} for further details.
We adopt the functional form for the halo mass function from Ref.~\cite{Tinker:2008ff},
\begin{equation}
    \frac{dn}{dM}=f\left(\sigma\right)\frac{\bar{\rho}_{m}}{M}\frac{d\ln\sigma^{-1}}{dM},\label{eq:hmf}
\end{equation}
where,
\begin{equation}
    \sigma^2(M,z)=\int_{k_\mathrm{min}}^{k_\mathrm{max}}\frac{dk}{2\pi^2} k^{2}P_{m}\left(k,z\right)W^2\left(kR\right),\label{eq:MassVariance}
\end{equation}
is the variance of the linear density field smoothed on scale $R\left(M\right)$, containing mass $M$; and
\begin{equation}
    f\left(\sigma\right)=A^\mathrm{T}\left[\left(\frac{\sigma}{b^\mathrm{T}}\right)^{-a^\mathrm{T}}+1\right]e^{-c^\mathrm{T}/\sigma^{2}},\label{eq:hmfUniFnction}
\end{equation}
is parameterized by $\left\{ A^\mathrm{T}, a^\mathrm{T}, b^\mathrm{T}, c^\mathrm{T} \right\}$, which depend on redshift, as summarized in Appendix~\ref{app:VidHMF}.\\
Therefore, modifications in the linear matter power spectrum, e.g.\ due to PMFs, affect both the galaxy number distribution and the abundance of DM halos at a given mass scale $M$, which alter the probability $\mathcal{P}_\mathrm{tot}(T)$, estimated via the histogram $B_i$.

\begin{figure*}[ht!]
    \centering
    \includegraphics[width=2\columnwidth]{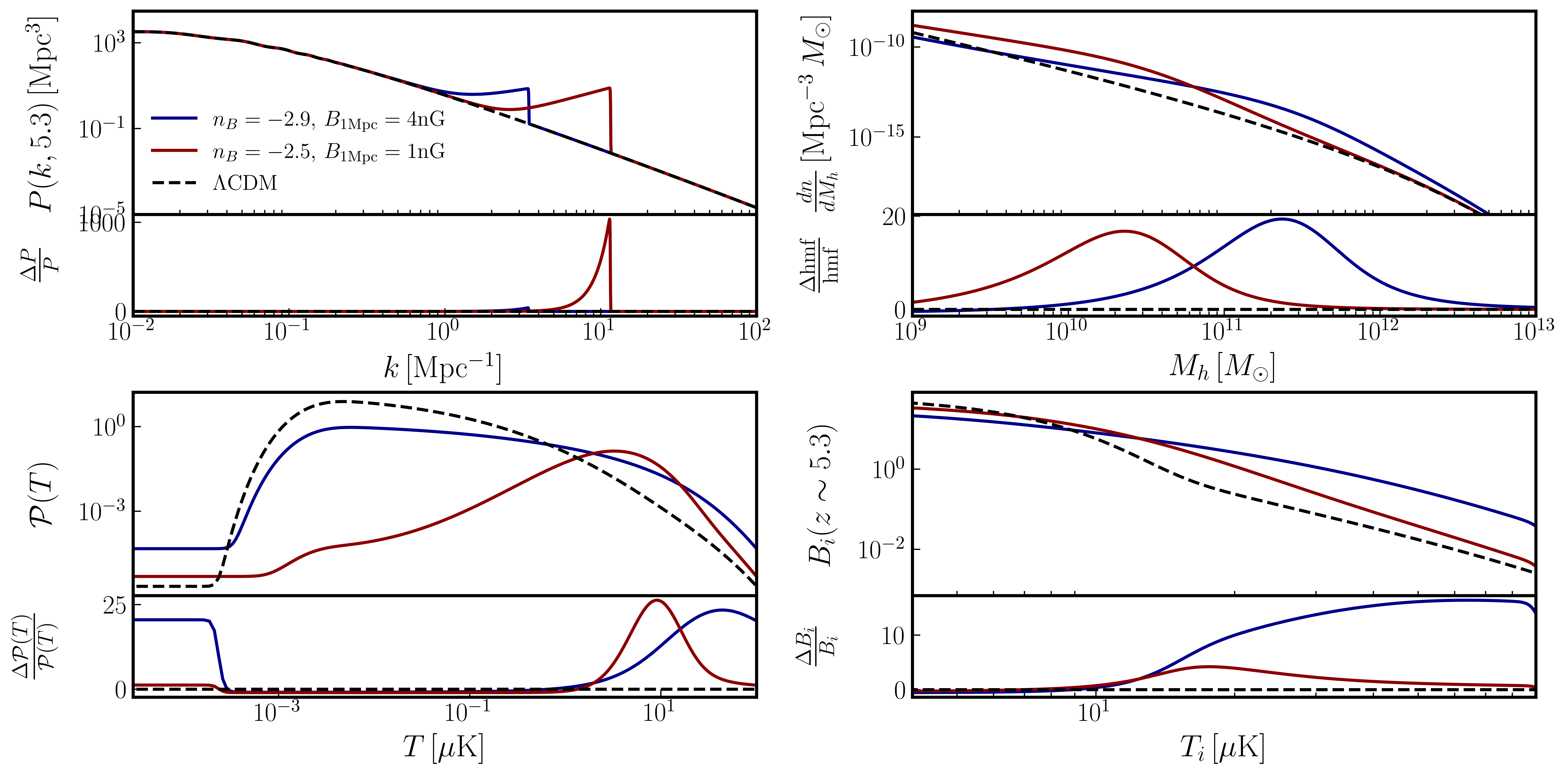}
    \caption{The VID pipeline for two sets of PMFs parameters (solid lines), compared to the $\Lambda$CDM scenario (dashed line), calculated at $z=5.3$: the linear matter power spectrum (top left); halo mass function (top right); the noise-free voxel temperature distribution (bottom left); and the VID, for the COS3 survey (bottom right). The enhanced power on small scales in the matter power spectrum leads to an abundance of faint sources, which boost the number of high-temperature bins.}
    \label{fig:VID_4panel}
\end{figure*}
\section{Analysis}
\label{sec:analysis}

Throughout this work we adopt the best-fit cosmological parameters from {\it Planck 2018} + BAO (right column of Table 2 in Ref.~\cite{Planck:2018vyg}).

\subsection{Magnetically Induced Matter Power Spectrum Computation}
In order to calculate the magnetically induced matter power spectrum in Eq.~\eqref{eq:Pm}, we evaluate the integral in Eq.~\eqref{eqn:PiPower} numerically, using the limits $k_\mathrm{min}<q< 3k_A$,\footnote{We take this upper limit to make sure we include all of the contributing scales.} where $k_\mathrm{min}=2\times10^{-4}\mathrm{Mpc}^{-1}$ is taken to be roughly the Hubble radius,\footnote{We checked that the resulting matter power spectrum is insensitive to lowering $k_\mathrm{min}$ by orders of magnitude.} and plugging in the PMFs power spectrum from Eq.~\eqref{eq:P_B-damped}. We calculate $k_A$ according to Eqs.~\eqref{eq:kAlfven}-\eqref{eq:vAlfven}, using the outputs from \texttt{CAMB}\footnote{\url{https://camb.info}}~\cite{Lewis:1999bs} to extract the ionization fraction history. The time evolution is calculated by solving Eq.~\eqref{eq:M-evo} numerically for $M(z)$, along with the standard background evolution of
\begin{equation}
    H^2(z) = H_0^2 \left[ \Omega_m\left(1+z\right)^3 + \Omega_r\left(1+z\right)^4 + \Omega_\Lambda\right],
\end{equation}
where $\Omega_i = \bar{\rho}_{i,0}/\rho_\mathrm{crit,0}$ and $ \rho_\mathrm{crit,0} = 3H_0^2/8\pi G$ are today's values of the dimensionless background energy densities and the critical density, respectively.

In order to exclude scales smaller than the magnetic Jeans scale, we introduce an exponential suppression function to the  magnetically induced matter power spectrum, $\mathcal{S}(k)=\exp{[-\left(k/k_\mathrm{crit} \right)^\kappa]}$, where $\kappa$ sets the suppression strength and $k_\mathrm{crit}=k_J\times0.01^{-\kappa}$ is defined to ensure that $\mathcal{S}(k_J)=0.99$; we set $\kappa=400$, which results in a very strong yet smooth enough suppression. In Figure~\ref{fig:pmf_vary} we show the contribution of the magnetically induced matter power spectrum on top of the $\Lambda$CDM linear matter power spectrum for several sets of values for the PMFs spectral index $n_B$ and strength $B_{\rm 1Mpc}\equiv B_{\rm 1Mpc}(a=1)$.

\subsection{VID Computation}\label{sec:vid_comp}

We compute the VID using a modified version of the public \texttt{LIM} code\footnote{\url{https://github.com/jl-bernal/lim}} described in Ref.~\cite{Bernal:2019jdo}. We use the halo mass function in Eqs.~\eqref{eq:hmf}-\eqref{eq:hmfUniFnction}, fixing its parameters, according to Ref.~\cite{Tinker:2008ff}, $\{A_0^\mathrm{T}, a_0^\mathrm{T}, b_0^\mathrm{T}, c^\mathrm{T}\} = \{0.186, 1.47, 2.57, 1.19\}$, and evolve them in time as described in Eq.~\eqref{eq:TinkerConsts}. Since the magnetically induced matter power spectrum features a sharp cut-off at the magnetic Jeans scale, using a real-space top-hat window function can lead to oscillatory behaviour in the mass variance and the halo mass function. This is a well-known issue and can be circumvented by using a sharp-k window function~\cite{Bertschinger:2006nq,Schneider:2013ria,Sabti:2021unj}, which we choose to be
\begin{equation}
    W(k) = \exp\left[-\frac{\left(kR/2\right)^2}{2}\right].
\end{equation}
The fiducial values we use for the astrophysical parameters of the CO luminosity relations in Appendix~\ref{app:VidLCO} are $\{\alpha, \beta, \sigma_{L_{\mathrm{CO}}}, \sigma_{\mathrm{SFR}}, L_\mathrm{cut}\}=\{1.37,-1.74, 0.3, 0.3, 50L_\odot\} $\footnote{We also tested our results with higher cutoff luminosity, $L_\mathrm{cut}=5000L_\odot$, and found no significant difference.}; taken from Ref.~\cite{Li:2015gqa}.
Our forecast focuses on two future surveys:
the planned COMAP-EoR (EoR) survey~\cite{COMAP:2021nrp}, equipped with a pair of instruments observing at $z\sim2-3$ and three instruments\footnote{As in Ref.~\cite{Libanore:2022ntl}, we do not consider the low frequency instrument observing CO from $z>7$, since astrophysical modelling in this range is still highly uncertain.} observing at $z\sim5-7$; and 
a stage-3 experiment (COS3, proposed in Ref.~\cite{Libanore:2022ntl}), with similar characteristics but a larger number of feeds, 
to increase the sensitivity. The specifications of both surveys, summarized in Table~\ref{tab:surveys}, affect the instrumental noise per voxel according to Eq.~\eqref{eq:sigmaNoise}.
In Figure~\ref{fig:VID_4panel}, we show how the magnetically enhanced matter power spectrum affects the halo mass function, which then propagates to the temperature fluctuations probability distribution $\mathcal{P}(T)$ and thus to the VID $B_i$.
\begin{table}[ht!]
\renewcommand{\arraystretch}{1.2}
    \centering
    \begin{tabular}{|c|c|c|c|c|c|c|}
    \hline
     \multirow{2}{*}{\parbox{1.5cm}{Survey}} & $\Delta\nu$ & $\Omega_\mathrm{field}$ & $\theta_{\rm FWHM}$ &$N_{\rm feeds}$ & $t_{\rm obs}$   & $\sigma_N$ \\
     & $[\mathrm{GHz}]$ & $[\mathrm{deg}^2]$ & $[\mathrm{arcmin}]$ & \# & $[\mathrm{hrs}]$ & $[\mu\mathrm{K}]$\\
    \hline
    \multirow{2}{*}{\parbox{2cm}{EoR / COS3\\low-$z$}} & 30\,-\,34 & 4  & 4.5' &  19 / $10^3$ & 
    $5\,000$ & 35 / 5.4    \\
     & 26\,-\,30 & 4  & 3.9' &  19 / $10^3$ & 
    $5\,000$ & 40 / 4.7 \\
    \hline
    \multirow{3}{*}{\parbox{1cm}{EoR\\high-$z$}} & 17\,-\,20 & 4  & 3.3' &  38 & 
    $7\,000$ & 22    \\
     & 15\,-\,17 & 4 & 3.7' & 38 &
    $7\,000$ & 16 \\
     & 13\,-\,15& 4& 4' & 38 &
    $7\,000\,$ & 14 \\
    \hline
    \multirow{3}{*}{\parbox{1cm}{COS3\\high-$z$}} & 17\,-\,20 &  4 & 3.3' & $10^3$ &
    $7\,000$ & 4.3\\
     & 15\,-\,17 & 4 & 3.7' & $10^3$  &
    $7\,000$ &   3.1 \\
     & 13\,-\,15 & 4 & 4' & $10^3$  & 
    $7\,000$ &  2.7 \\
    \hline
    \end{tabular}
    \caption{Summary of the detectors specifications: frequency range $\Delta\nu$, field of view $\Omega_\mathrm{field}$, angular resolution $\theta_{\rm FWHM}$, number of feeds $N_\mathrm{feeds}$, observational time $t_{\rm obs}$.}
    \label{tab:surveys}
\end{table}
\subsection{Fisher Analysis}
We perform a Fisher analysis to evaluate the sensitivity of the VID of the surveys under consideration to changes induced by PMFs.
Following Refs.~\cite{Breysse:2015saa,Libanore:2022ntl}, we use
\begin{equation}\label{eq:fisher}
    F_{\alpha\beta} = \sum_D\sum_{i>T_i^\mathrm{min}} \frac{1}{\sigma_i^2} \frac{d\left(N_zB_i\right)}{d\theta_\alpha} \frac{d\left(N_zB_i\right)}{d\theta_\beta},
\end{equation}
where $D$ is the number of detectors of the survey, $i$ denotes the $i$-th temperature bin, $\sigma_i$ is the variance per bin and $N_z=\Delta\nu/\delta\nu$ is the number of frequency channels, of width $\delta\nu=2\,\mathrm{MHz}$, in each frequency band. We assume the signal does not evolve within the redshift slices spanned by the bandwidth. We also assume Poisson uncertainties for the number of pixels in each bin, so that $\sigma_i^2=N_zB_i$~\cite{Breysse:2014uia}.
The Fisher matrix in Eq.~\eqref{eq:fisher} is computed for the PMFs and astrophysical parameters, namely $\theta=\{n_B, B_{\rm 1Mpc}; \alpha, \beta, \sigma_\mathrm{SFR}, \sigma_{L_\mathrm{CO}}, L_\mathrm{cut}\}$, using the fiducial values mentioned in Sec.~\ref{sec:analysis} while varying the fiducial values of the PMFs parameters.\footnote{We limit our analysis to $-2.99\le n_B\le -1$, also covering inflationary originated PMFs~\cite{PhysRevD.37.2743,Turner:1987bw,Ratra:1991bn}, which seems preferable by observational evidence~\cite{Caprini:2001nb,Sethi:2004pe,Jedamzik:2018itu}.} Cosmological parameters, as well as halo mass function parameters are held fixed throughout this work: the former are well measured by {\it Planck}~\cite{Planck:2018vyg} and have negligible errorbars with respect to other quantities in the analysis, while the effect of the latter on the final results is subdominant~\cite{Libanore:2022ntl}.
\section{Results}
\label{sec:results}
\begin{figure}[ht!]
    \centering
    \includegraphics[width=\columnwidth]{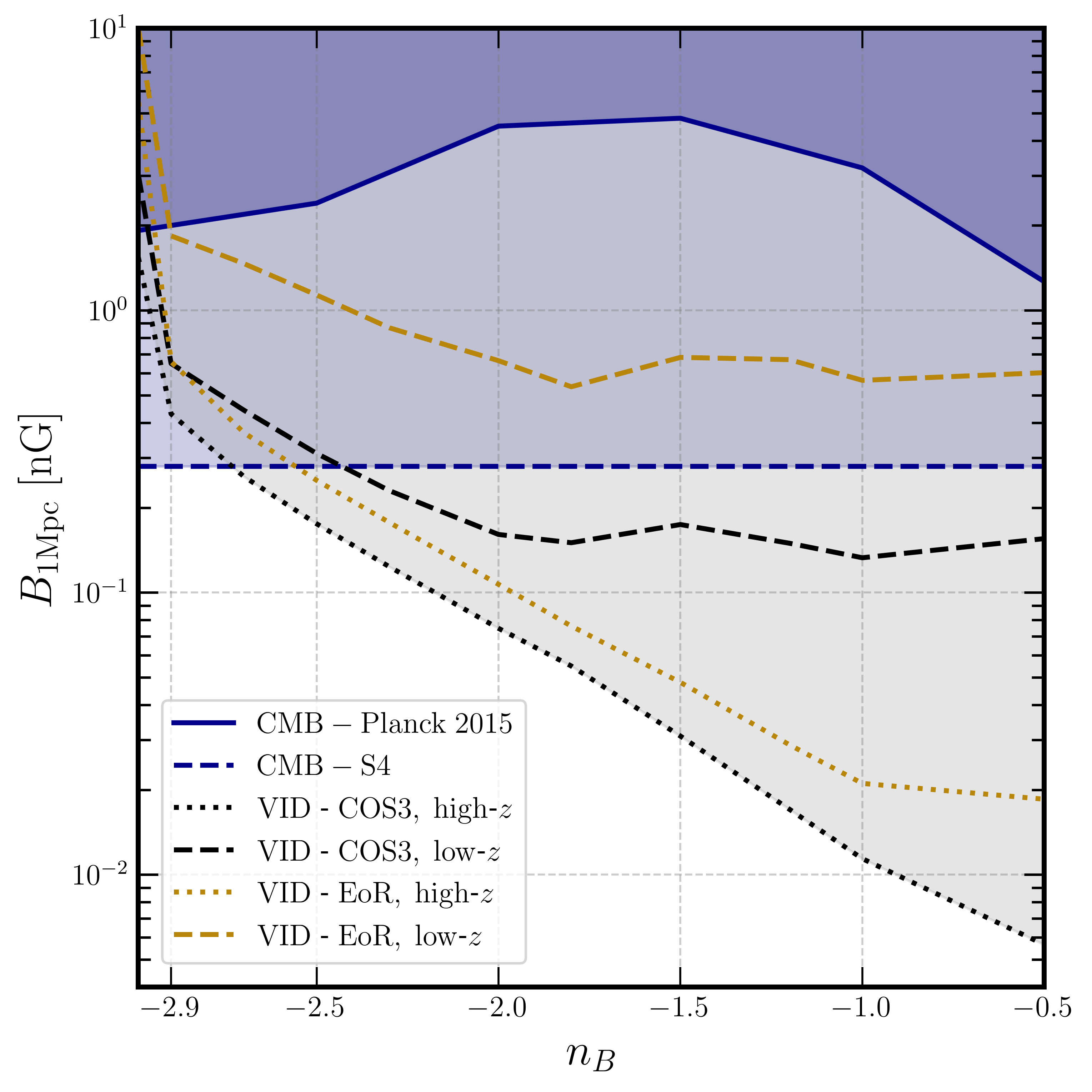}
    \caption{Summary of the bounds (solid lines) and sensitivity forecasts (dashed and dotted lines) for the strength of PMFs at 95\% CL. We show forecast for the VID from different CO LIM surveys along with forecast for CMB-S4 experiment~\cite{Sutton:2017jgr}; compared to the bounds from CMB, measured by {\it Planck 2015}~\cite{Planck:2015zrl}.}
    \label{fig:bounds_summary}
\end{figure}

Our main results for the forecast of VID sensitivity for COS3 and EoR LIM surveys are summarized in Table~\ref{tab:bounds_summary} and Figure~\ref{fig:bounds_summary}. We include recent bounds\footnote{Other bounds, e.g., from reionization history~\cite{Pandey:2014vga} and Ly$\alpha$ clouds~\cite{Pandey:2012ss} exist in the literature. However, these miss the $(4\pi)^2$ factor in Eq.~\eqref{eqn:PiPower} and therefore should be considered with care, since they are expected to worsen if analyzed properly.} from CMB measurements by {\it Planck 2015} (TT,TE,EE+lowP)~\cite{Planck:2015zrl}, as well as a forecast for CMB-S4~\cite{Sutton:2017jgr} as a reference. We note that other CMB forecasts exist in the literature, e.g., from anisotropic birefringence measurements~\cite{Mandal:2022tqu}, as well as non-observational bounds, e.g., from reionization, captured by detailed radiation-hydrodynamical simulations~\cite{Katz:2021iou}.

The excess of power on small scales, due to the magnetically induced matter power spectrum, leads to an abundance of lower-mass halos, as shown in Figure~\ref{fig:VID_4panel}. 
This translates to an increased number of faint sources, hence more high-temperature bins. Since the Gaussian noise dominates the low-temperature bins, the large number of faint sources yields a relatively clean signal of the VID.\footnote{One should keep in mind that other deviations from the $\Lambda$CDM model may give rise to suppression of the signal, e.g., due to fuzzy dark matter~\cite{Flitter:2022pzf, Hlozek:2016lzm, Dentler:2021zij}.}

\begin{table}[htbp]
\renewcommand{\arraystretch}{1.2}
    \centering
    \begin{tabular}{|c|c|ccccc|}
    \hline
    \multicolumn{2}{|c|}{$n_B$} & -1 & -1.5 & -2 & -2.5 & -2.9 \\
    \specialrule{.1em}{.05em}{.05em}
    \multirow{4}{*}{\parbox{2cm}{$B_{\rm 1Mpc}\left[\mathrm{nG}\right]$}}
    & Planck 2015 & 3.2 & 4.8 & 4.5 & 2.4 & 2.0 \\
    \cline{2-7}
    & CMB-S4 & \multicolumn{5}{c|}{$\sim0.28$}\\
    \cline{2-7}
    & EoR loz-$z$ & 0.565 & 0.682 & 0.663 & 1.134 & 1.838 \\
    \cline{2-7}
    & COS3 high-$z$ & 0.011 & 0.031 & 0.075 & 0.175 & 0.43 \\
    \hline
    \end{tabular}
    \caption{The {\it Planck 2015}~\cite{Planck:2015zrl} bounds on PMFs strength for chosen values of $n_B$, compared to the sensitivity forecast for CMB-S4~\cite{Sutton:2017jgr} and the VID of most and least sensitive surveys we considered, all at 95\% CL.}
    \label{tab:bounds_summary}
\end{table}

The results for the low-$z$ surveys, shown in Figure~\ref{fig:bounds_summary}, suggest that the VID sensitivity deteriorates for values of $n_B\gtrsim-1.5$. The reason for this lies mainly in the way the PMFs parameters, $n_B$ and $B_{\rm 1Mpc}$, affect the magnetic Jeans scale. From Figure~\ref{fig:pmf_vary}, it is easy to see that reducing either parameter results in a larger $k_J$, but not to the same extent. For a given value of $n_B$, reducing $B_{\rm 1Mpc}$ also means shifting the excess of power to smaller scales, which translate to smaller halos that host fainter sources. These faint sources give rise to low temperature voxels which are noise dominated, thus the making the PMFs contribution to matter overdensities harder to detect by LIM. This effect is stronger for lower redshifts, at which larger halos are dominant. In addition, scales corresponding to masses near the infrared mass-cutoff at $\sim10^9M_\odot$ do not contribute to the LIM signal.

Now, consider a fixed value for $B_{\rm 1Mpc}$. By going to lower $n_B$ the power shifts to smaller scales, leading to weaker signal, as described above. While increasing $n_B$ enhances larger scales, which naively result in more luminous sources, the halo mass function suppresses the number density of massive halos. Therefore, we expect the LIM sensitivity to decline above some threshold $n_{B}^\mathrm{max}$. This effect becomes more pronounced at lower redshifts.

We note that our analysis is limited to the simplest, first order, effect of PMFs on halo formation. For instance, the dissipation of PMFs is expected to heat the IGM, as was suggested in Refs.~\cite{Sethi:2004pe, Sethi:2008eq}, which in turn could hinder structure formation and alter the time and scales at which stars and galaxies are formed~\cite{Tashiro:2005ua, Sanati:2020oay}. Still, for the spectral indices and amplitudes we consider in our analysis, these effects are not expected to have a significant impact. We leave the inclusion of PMFs heating effects to future work.

\section{Conclusions}
\label{sec:conclusions}

In this work we studied the sensitivity of ongoing and next generation CO LIM surveys~\cite{COMAP:2021nrp,Libanore:2022ntl} to the signature of PMFs in the matter power spectrum~\cite{Kim:1994zh} at small scales $k\sim1-10^{2}\,\mathrm{Mpc}^{-1}$, using the one-point correlation of the voxel temperature distribution, commonly referred to as the VID. Recent work showed that the VID in intensity maps is sensitive to smaller scales than those accessed by the surveys angular resolution~\cite{Libanore:2022ntl}, making LIM an interesting probe of the magnetically induced matter overdensities. The expression for the magnetic contribution to the linear matter power spectrum in Eq.~\eqref{eqn:PiPower}, that is derived in Ref.~\cite{Adi:4pi}, is smaller by a factor of $(4\pi)^2$ than the one that was used in previous studies, e.g., Refs.~\cite{Pandey:2012ss,Pandey:2014vga}. This difference affects the bounds in a non-trivial way. Our results, summarized in Table~\ref{tab:bounds_summary} and Figure~\ref{fig:bounds_summary}, show that the VID of CO line intensity maps are expected to be sensitive to magnetic field strengths according to $\log_{10}\left(B_{\rm 1Mpc}/1\mathrm{nG}\right)\gtrsim -(2.7+0.8n_B)$ for  power-law PMFs power spectra with spectral index $-2.9\lesssim n_B\lesssim -0.5$. With such sensitivity, LIM stands to outdo next-generation CMB experiments.

\acknowledgements

We thank Marc Kamionkowski and Jordan Flitter for useful discussions and comments. TA~is supported by a Negev PhD fellowship awarded by the BGU Kreitmann School.
SL~acknowledges the Azrieli Foundation for the support. HAC is supported by the National Science Foundation Graduate Research Fellowship under Grant No.\ DGE2139757. EDK~acknowledges support from an Azrieli faculty fellowship.

\appendix
\setcounter{footnote}{0}

\section{Evolution of Matter Fluctuations}
\label{app:PMF}
In what follows, we will use the standard notations: gravitational potential $\psi=\phi$, background densities $\bar{\rho}_i$, density contrasts $\delta\rho_i=\rho_i/\bar{\rho}_i-1$, fluid velocities $v_i$, with $i=b,c$ for baryons and cold dark matter (CDM).\\
After recombination, any PMFs that were frozen in the baryon-photon plasma are 
released and obey\footnote{We drop the $\nabla^2\bs{B}/\mu_0\sigma$ term, as it is negligible for high electric conductivity and scales larger than the magnetic Jeans scale.}
\begin{align}
    \bs{\nabla}\cdot\bs{B} &= 0\,,\\
    \frac{\partial}{\partial t}\left(a^2 \bs{B}\right) &= \frac{\bs{\nabla}\times\left(\bs{v}\times a^2\bs{B}\right)}{a}\,.\label{eq:lorentz}
\end{align}
These magnetic fields add an additional source term in the linearized equations for the evolution of baryon density contrast, which obeys the fluid equations
\begin{align}
    \dot{\bs{v}}_b + H \bs{v}_b &= -\frac{\bs{\nabla}\psi}{a} + \frac{\left( \bs{\nabla}\times\bs{B} \right) \times \bs{B}}{\mu_0 a},\label{eq:euler-b}\\
    \dot{\bs{v}}_c + H \bs{v}_c &= -\frac{\bs{\nabla}\psi}{a},\\
    \dot{\delta}_i+\frac{\bs{\nabla}\cdot\bs{v}}{a} &= 0,\\
    \dot{\bar{\rho}}_i + 3 H \bar{\rho}_i &= 0,
\end{align}
where $\mu_0$ is the vacuum magnetic permeability. Along with the Poisson equation
 \begin{equation}
     \nabla^2\psi = 4\pi G a^2\bar{\rho}_m\delta_m\,,
 \end{equation}
where $\bar{\rho}_m\delta_m \equiv \bar{\rho}_b\delta_b + \bar{\rho}_c\delta_c$, one finds that\footnote{Note that the source term had pick up a sign, resulting in the reordering of the multiplicands. As this subtlety eluded some authors, one should be careful when using the literature.}
\begin{equation}
    \ddot{\delta}_m+2H\dot{\delta}_m - 4\pi G\bar{\rho}_m\delta_m = \frac{\bar{\rho}_{b,0}}{\bar{\rho}_{m,0}} \frac{\bs{\nabla}\cdot\left[ \bs{B} \times \left(\bs{\nabla}\times \bs{B}\right) \right]}{\mu_0\bar{\rho}_ba^2},\label{eq:delta_m-eq1}
\end{equation} 
 with subscript $0$ denoting the value of the quantity today.\\
For scales much larger than the magnetic Jeans scale, the term on the right hand side of Eq.~\eqref{eq:lorentz} vanishes, which sets the time evolution of magnetic fields on large scales as $\bs{B}(\bs{x},t) = \bs{B}_0/a^2$. Plugging this relation into Eq.~\eqref{eq:delta_m-eq1}, together with the background evolution of the baryon density $\bar{\rho}_b = \bar{\rho}_{b,0}/a^3$, yields
\begin{equation}
    \ddot{\delta}_m + 2H\dot{\delta}_m - 4\pi G\bar{\rho}_m\delta_m = f_b \frac{\bs{\nabla}\cdot\left[ \bs{B}_0 \times \left(\bs{\nabla}\times \bs{B}_0\right) \right]}{\mu_0\bar{\rho}_{b,0}a^3},\label{eq:delta_m-eq2}
\end{equation}
where we defined $f_b \equiv \bar{\rho}_b / \bar{\rho}_m$. The homogeneous solution to this equation leads to the standard linear evolution of the matter density contrast, which can be written as $\delta_m^\mathrm{lin}(t) = A_+(\bs{x}) D_+(t) + A_-(\bs{x}) D_-(t)$, where $\pm$ denote the growing and decaying solutions. Whereas the special solution can be written as
\begin{align}
    \delta_m^\mathrm{mag}(t) &= f_b M(t) \bs{\nabla}\cdot\bs{S}_0(\bs{x}),\label{eq:delta_mB-sol}\\
    \bs{S}_0(\bs{x}) &\equiv \frac{ \bs{B}_0 \times \left(\bs{\nabla}\times \bs{B}_0\right) }{\mu_0\bar{\rho}_{b,0}},
\end{align}
so that in total
\begin{equation}
    \delta_m(t) = \delta_m^\mathrm{lin}(t) + \delta_m^\mathrm{mag}(t).\label{eq:delta_m2}
\end{equation}
Plugging Eq.~\eqref{eq:delta_mB-sol} into Eq.~\eqref{eq:delta_m-eq2} yields the equation for the \textit{magnetic growth function} $M(t)$,
\begin{align}
    \ddot{M}+2H\dot{M}-4\pi G\bar{\rho}_{m,0}\frac{M}{a^3} &= \frac{1}{a^3},\\
    M(t_\mathrm{rec})=\dot{M}(t_\mathrm{rec}) &= 0.\nonumber
\end{align}
Therefore, assuming no correlation between the standard and the magnetically induced matter density contrasts, the modified matter power spectrum follows Eq.~\eqref{eq:Pm}, i.e.,
\begin{equation}
    P_m(k,t) = D_+^2(t)P_\mathrm{lin}(k) + M^2(t)\Pi(k),
\end{equation}
where
\begin{equation}
    \Pi(k) = \left| \frac{f_{\mathrm{b}}}{\mu_0 \rho_{\mathrm{b}, 0}} \int d^3 y\; \bs{\nabla}\cdot\left[\bs{B}_0 \times \left(\bs{\nabla}\times \bs{B}_0\right)\right]e^{-i\bs{k}\cdot\bs{y}} \right|^2 \label{eq:PiPower-general}
\end{equation}
is the scale dependent term, which after a careful calculation~\cite{Kim:1994zh} yields Eq.~\eqref{eqn:PiPower}.
\section{VID Formalism}
\label{app:Vid}
In this Appendix we briefly summarize the VID formalism used throughout this work, following Refs.~\cite{Breysse:2014uia,Breysse:2016szq,Bernal:2020lkd,Libanore:2022ntl}.

\subsection{Analytical Modeling of the VID}
\label{app:VidModeling}

The luminosity density of a given emission line, e.g., CO in the main text, denoted by $L_{\rm CO} = L$, can be written in terms of the luminosity or mass function as
\begin{equation}
    \rho_{L}=\int_{0}^{\infty}L\frac{dn}{dL}dL=\int_{0}^{\infty}L\left(M\right)\frac{dn}{dM}dM,
\end{equation}
so that the relation between the local luminosity density and brightness temperature is given by\footnote{This can be derived directly from the Rayleigh-Jeans relation, see for example Ref.~\cite{Bernal:2022jap}.}
\begin{equation}
    T\left(\boldsymbol{x}\right)=\frac{c^{3}\left(1+z\right)^{2}}{8\pi k_{B}\nu^{3}_\mathrm{rest}H\left(z\right)}\rho_{L}\left(\boldsymbol{x}\right)\equiv X_{\mathrm{LT}}\left(z\right)\rho_{L}\left(\boldsymbol{x}\right),
\end{equation}
where $\nu_\mathrm{rest}$ is the rest-frame frequency of the line.
The VID is the histogram of the measured signal within the voxels. We estimate the brightness temperature inside a voxel of volume $V_{\mathrm{vox}}$ that contains $N_s$ sources as
\begin{equation}
    T=\frac{X_{\mathrm{LT}}}{V_{\mathrm{vox}}}\sum_{i=1}^{N_s}L_{i},
\end{equation}
where $L_{i}$ is the luminosity of the $i$-th source. The probability of observing temperature $T$ for a voxel with a single source is equivalent to the probability that a source has the corresponding luminosity $L_{T}=TV_{\mathrm{vox}}/X_{\mathrm{LT}}$, namely 
\begin{equation}
    \mathcal{P}_{1}\left(T\right)=\frac{V_{\mathrm{vox}}}{\bar{n}X_{\mathrm{LT}}}\frac{dn}{dL}\big|_{L_{T}},
\end{equation}
where $\bar{n}$ normalizes the luminosity function, and the factor $V_{\mathrm{vox}}/X_{\mathrm{LT}}$ converts $L$ into brightness temperature $T$. Considering a voxel with two sources, such that the brightness temperature of the voxel is $T=T_{a}+T_{b}$, the probability to find brightness temperature $T$ is
\begin{equation}
    \mathcal{P}_{2}\left(T\right)=\int dT_{a}\mathcal{P}_{1}\left(T_{a}\right)\mathcal{P}_{1}\left(T-T_{a}\right)=\left[\mathcal{P}_{1}*\mathcal{P}_{1}\right]\left(T\right),
\end{equation}
which is the convolution of the two single source probabilities. Iterating this argument for $N_{s}$ sources, one finds that the probability for a voxel with $N_{s}$ sources to have a brightness temperature $T$ is
\begin{equation}
    \mathcal{P}_{N_{s}}\left(T\right)=\left[\mathcal{P}_{N_{s}-1}*\mathcal{P}_{1}\right]\left(T\right).
\end{equation}
Thus, the probability of having a voxel with brightness temperature $T$ is simply
\begin{equation}
    \mathcal{P}\left(T\right) = \sum_{N_{s}=0}^{\infty} \mathcal{P}_{N_{s}} \left(T\right) \mathcal{P}_{s}\left(N_{s}\right),
\end{equation}
where $\mathcal{P}_{s}\left(N_{s}\right)$ is the probability of having $N_{s}$ sources in a voxel.
To model a future observation in a semi-analytical way, we convolve the signal $\mathcal{P}\left(T\right)$ with the instrumental noise per voxel, which we assume to follow a Gaussian distribution~\cite{Breysse:2016szq} with zero mean, and variance
\begin{equation}\label{eq:sigmaNoise}
    \sigma_{N}=\sqrt{\frac{P_{\mathrm{N}}}{V_{\mathrm{vox}}}}=T_{\mathrm{sys}}\sqrt{\frac{N_{\mathrm{vox}}}{N_{\mathrm{feeds}}t_{\mathrm{obs}}\delta\nu}},
\end{equation}
where $P_{\mathrm{N}}$ is the noise power spectrum, $T_{\mathrm{sys}}$ is the system temperature, $t_{\mathrm{obs}}$ is the observation time, $N_{\mathrm{vox}}$ and $N_{\mathrm{feeds}}$ are the number of voxels and feeds respectively, and $\delta\nu$ is the frequency channel width. In this way 
\begin{align}
    \mathcal{P}_{n}\left(T\right)&=\frac{1}{\sqrt{2\pi}\sigma_{N}}\exp\left[-\frac{T^{2}}{2\sigma_{N}^{2}}\right],\\
    \mathcal{P}_{\mathrm{tot}}\left(T\right)&=\left(\mathcal{P}*\mathcal{P}_{n}\right)\left(T\right),
\end{align}
Finally, the VID is 
the histogram of the number of voxels $B_{i}$ for which the measured brightness temperature is found within a bin $\Delta T_{i}$, such that
\begin{equation}
    B_{i}=N_{\mathrm{vox}}\int_{T_{i}}^{T_{i}+\Delta T_{i}}\mathcal{P}_{\mathrm{tot}}\left(T\right)dT.
\end{equation}
So basically, in order to infer the VID, all we need is to provide three relations: the halo mass function $dn/dM$, the line luminosity for a given halo mass $L\left(M\right)$, and the luminosity function $dn/dL$. Since the latter can be, naively, inferred from the former two, it would be beneficial to use such relation.
\subsection{The Halo Mass Function}
\label{app:VidHMF}
The halo mass function in Eqs.~\eqref{eq:hmf}-\eqref{eq:hmfUniFnction}, adopted from Ref.~\cite{Tinker:2008ff}, is parameterized by $\left\{ A^\mathrm{T},a^\mathrm{T},b^\mathrm{T},c^\mathrm{T}\right\}$, which are calibrated via simulations, typically at $z=0$. The parameter $A^\mathrm{T}$ sets the overall amplitude of the mass function, $a^\mathrm{T}$ and $b^\mathrm{T}$ determine the slope and amplitude of the low mass power law, while $c^\mathrm{T}$ sets the cutoff scale for the exponential decrease in the halo abundance. These parameters are found to vary with redshift according to 
\begin{align}
    A^\mathrm{T}\left(z\right)	&=A_{0}^\mathrm{T}\left(1+z\right)^{-0.14},\nonumber\\
    a^\mathrm{T}\left(z\right)	&=a_{0}^\mathrm{T}\left(1+z\right)^{-0.06},\nonumber\\
    b^\mathrm{T}\left(z\right)	&=b_{0}^\mathrm{T}\left(1+z\right)^{-\alpha^\mathrm{T}},\label{eq:TinkerConsts}\\
    \log\alpha^\mathrm{T}\left(\Delta\right)	&=-\left[\frac{0.75}{\log\Delta/75}\right]^{1.2},\nonumber
\end{align}
where
\begin{equation}
    \Delta\equiv\frac{M}{\left(4\pi/3\right)R^{3}}/\bar{\rho}_{m}\left(z\right).
\end{equation}
The values and functional forms we report are the fiducial results of the analysis in Ref.~\cite{Tinker:2008ff}; check the discussion there for uncertainties and limits in these relations.
\subsection{CO Luminosity - Halo Mass Relation}
\label{app:VidLCO}
The CO luminosity is derived from a series of relations, as described in Ref.~\cite{Li:2015gqa}:
\begin{enumerate}
    \item DM halos to star formation rate (SFR), inferred from simulation results
    ~\cite{2013ApJ...770...57B,2013ApJ...762..109B} via interpolation of
    $\mathrm{SFR}\left(M,z\right)$ with a lognormal scatter, parameterized by $\sigma_{\mathrm{SFR}}=0.3$.
    \item SFR to infrared (IR) luminosity,
            \begin{equation}
                L_{\mathrm{IR}}=\frac{\mathrm{SFR}\left(M,z\right)}{10^{-10}\delta_{\mathrm{MF}}},
            \end{equation}
            where $\mathrm{SFR}$ is in $\mathrm{M}_{\odot}\mathrm{yr}^{-1}$ and $L_{\mathrm{IR}}$ is in units of $L_{\odot}$. The normalization $0.8\lesssim\delta_{\mathrm{MF}}\lesssim2$ can be absorbed into the parameters of the $L_{\mathrm{CO}}$ to $L_{\mathrm{IR}}$ relation below; for the other parameters we adopt the fiducial values from Ref.~\cite{Carilli:2013qm}, $\left\{ \alpha=1.37,\beta=-1.74\right\}$, adding a lognormal scatter $\sigma_{\mathrm{IR}}=0.3$ to $L_{\mathrm{CO}}$, for a given $L_{\mathrm{IR}}$ or $\mathrm{SFR}$, which is consistent with previous studies~\cite{Daddi:2009pc,Carilli:2013qm,2014ApJ...794..142G,refId0}.
    \item CO luminosity,
            \begin{equation}
                L_{\mathrm{CO}}=\left[10^{-\beta}L_{\mathrm{IR}}\right]^{1/\alpha},
            \end{equation}
            where $L_{\mathrm{IR}}$ is in units of $L_{\odot}$, $L'_{\mathrm{CO}}$ is in units of $\left(\mathrm{K}\,\mathrm{km}\,\mathrm{s}^{-1}\mathrm{pc}^{2}\right)$ (observer units for velocity- and area-integrated brightness temperature), and
            \begin{equation}
                L_{\mathrm{CO}}\propto\left(\frac{L'_{\mathrm{CO}}}{\mathrm{K}\,\mathrm{km}\,\mathrm{s}^{-1}\mathrm{pc}^{2}}\right)\left(\frac{\nu_{\mathrm{rest,CO}}}{\nu_{\mathrm{rest,CO}_{1\rightarrow0}}}\right)^{3}L_{\odot},
            \end{equation}
            with $\nu_{\mathrm{rest,CO}}/\nu_{\mathrm{rest,CO}_{1\to 0}}$ the ratio between a CO rotational transition line and its ground state transition calculated in the source rest frame, i.e., $\nu_{\mathrm{rest,CO}_{1\rightarrow0}} = 115.271\,{\rm GHz}$. In our analysis we only consider the $1\to 0$ transition, so $L_{\rm CO}'\to L_{\rm CO}$ only accounts for units conversion.
\end{enumerate}
\subsection{The Luminosity Function}
\label{app:VidLF}
In order to compute $\mathcal{P}_{1}\left(T\right)$ we use the luminosity function, which, naively, can be written in terms of the halo mass function by using the $L\left(M\right)$ relation, as
\begin{equation}\label{eq:dndl_dndm}
    \frac{dn}{dL}=\frac{dn}{dM}\left(\frac{dL}{dM}\right)^{-1}.
\end{equation}
Since more than one halo mass value can produce the same luminosity, e.g., due to the scatter in the $L\left(M\right)$ relation, the relation in Eq.~\eqref{eq:dndl_dndm} is no monotonous and one should use another method for evaluating the luminosity function from the halo mass function. To overcome this issue, we use the conditional probability distribution $\mathcal{P}\left(L|M\right)$ of having a luminosity $L$ coming from a halo of mass $M$, such that
\begin{equation}
    \frac{dn}{dL}=\int dM\mathcal{P}\left(L|M\right)\frac{dn}{dM},
\end{equation}
where
\begin{equation}
    \mathcal{P}\left(L|M\right)=\frac{e^{-L_{\mathrm{cut}}/L}}{\sqrt{2\pi}L\sigma_{\mathrm{LN}}}\exp\left\{ -\frac{\left(\ln L-\mu_{\mathrm{LN}}\right)^{2}}{2\sigma_{\mathrm{LN}}^{2}}\right\}, 
\end{equation}
\begin{equation*}
    \mu_{\mathrm{LN}}=\ln\left[\frac{\mu_{L}^{2}}{\sqrt{\mu_{L}^{2}+\sigma_{L}^{2}}}\right];\quad \sigma_{\mathrm{LN}}^{2}=\ln\left(1+\frac{\sigma_{L}^{2}}{L_{\mathrm{CO}}^{2}}\right).
\end{equation*}
We added an exponential cutoff to reflect the luminosity cutoff at the low mass end of the luminosity function. Taking the mean for the CO luminosity as $\mu_{L}=L_{\mathrm{CO}}\left(M\right)$ and the lognormal variance to account for the scatter in both relations $\sigma_{\mathrm{LN}}^{2}\equiv\sigma_{\mathrm{TOT}}^{2}=\sigma_{L_{\mathrm{CO}}}^{2}+\sigma_{\mathrm{SFR}}^{2}/\alpha^{2}$, in which the $1/\alpha^{2}$ factor comes from the $L_{\mathrm{IR}}-L_{\mathrm{CO}}$ relation, we find that
\begin{equation}
    \frac{\sigma_{L}^{2}}{L_{\mathrm{CO}}^{2}}=e^{\sigma_{\mathrm{TOT}}^{2}}-1; \quad \mu_{\mathrm{LN}}=\ln L_{\mathrm{CO}}-\frac{1}{2}\sigma_{\mathrm{TOT}}^{2},
\end{equation}
so that
\begin{align}
    \mathcal{P}&\left(L|M\right)=\frac{e^{-L_{\mathrm{cut}}/L}}{\sqrt{2\pi}L\sigma_{\mathrm{TOT}}}\times\\
    \times&\exp\left\{ -\frac{\left[\left(\ln L-\ln\left[L_{\mathrm{CO}}\left(M\right)\right]\right)+\sigma_{\mathrm{TOT}}^{2}/2\right]^{2}}{2\sigma_{\mathrm{TOT}}^{2}}\right\},\nonumber
\end{align}
leading to Eq.~\eqref{eq:dndl}.

\subsection{The Source Distribution}
\label{app:VidSourceDist}
CO line emission is sourced in star forming galaxies; to draw their observed number within a voxel, we associate it a lognormal-distributed~\cite{Coles:1991if} expectation number of galaxies $\mu$, which in turn serves as the mean for a Poisson distribution
\begin{equation}
    \mathcal{P}_{s}\left(N_{s}\right) = \int_{0}^{\infty}\mathcal{P}_{LN} \left(\mu\right) \mathcal{P}_{\mathrm{Poiss}}\left(N,\mu\right)d\mu.
\end{equation}
Instead of $\mu$, one may consider the galaxy density field contrast~\cite{2002aprm.conf..299T}, $\delta_{LN}=\frac{\mu}{\overline{N}}-1$, where $\overline{N}$ is the mean number of galaxies, which can be written in terms of a Gaussian field $\delta_{G}$ as
\begin{equation}
    1+\delta_{LN}=\frac{\mu}{\overline{N}}=e^{\delta_{G}-\sigma_{G}^{2}/2},
\end{equation}
where $\sigma_{G}$ is the variance of the field, so that
\begin{equation}
    \mathcal{P}_{LN}\left(\mu\right)=\frac{1}{\mu\sqrt{2\pi\sigma_{G}^{2}}}\exp\left\{ -\frac{1}{2\sigma_{G}^{2}}\left[\ln\left(\frac{\mu}{\overline{N}}\right)+\frac{\sigma_{G}^{2}}{2}\right]^{2}\right\} .
\end{equation}
In order to calculate $\sigma_{G}$, we use the power spectrum of the Gaussian field $P_{G}\left(k\right)$ and the voxel's window function $W\left(\boldsymbol{k}\right)$, hence
\begin{equation}
    \sigma_{G}=\int P_{G}\left(k\right)\left|W\left(\boldsymbol{k}\right)\right|^{2}\frac{d^{3}\boldsymbol{k}}{\left(2\pi\right)^{3}},
\end{equation}
where $P_{G}\left(k\right)$ is calculated from the biased matter power spectrum $P_{LN}\left(k\right)=\overline{b}^{2}P_{m}\left(k\right)$, by taking advantage of the relation between the real space correlation functions~\cite{Coles:1991if} $\xi_{G}\left(\boldsymbol{r}\right)=\ln\left[1+\xi_{LN}\left(\boldsymbol{r}\right)\right]$.Whereas the marginalized bias is given by
\begin{equation}
    \overline{b}\left(z\right)=\frac{\int Lb\left(L\right)\frac{dn}{dL}dL}{\int L\frac{dn}{dL}dL}=\frac{\int L\left(M\right)b\left(M\right)\frac{dn}{dM}dM}{\int L\left(M\right)\frac{dn}{dM}dM},
\end{equation}
where $b\left(M\right)$ is the mass-dependent halo bias~\cite{Breysse:2014uia},
\begin{equation}
    b\left(M,z\right)=1+\frac{\nu\left(M,z\right)-1}{\delta_{c}},
\end{equation}
with $\delta_{c}=1.69$, $\nu\left(M,z\right)=\delta_{c}/\sigma\left(M,z\right)$, and $\sigma\left(M,z\right)$ is the rms mass variance.
\bibliography{refs}

\end{document}